\input psfig
\newcommand {\sax} {{\it BeppoSAX }}

\def \rsun {\ifmmode$R$_{\odot}\else R$_{\odot}$\fi}

\def \hcm {\hbox {\ifmmode $ H atoms cm$^{-2}\else H atoms cm$^{-2}$\fi}}
\def\approxgt{\mathrel{\hbox{\rlap{\lower.55ex \hbox {$\sim$}}
        \kern-.3em \raise.4ex \hbox{$>$}}}}
\def\approxlt{\mathrel{\hbox{\rlap{\lower.55ex \hbox {$\sim$}}
        \kern-.3em \raise.4ex \hbox{$<$}}}}
\def\la{\mathrel{\hbox{\rlap{\hbox{\lower4pt\hbox{$\sim$}}}\hbox{$<$}}}}
\def\ga{\mathrel{\hbox{\rlap{\hbox{\lower4pt\hbox{$\sim$}}}\hbox{$>$}}}}
\def\kms{${\rm km~s^{-1}}$}

\documentstyle[twoside,fleqn,espcrc2]{article}

\newcommand{\AmS}{{\protect\the\textfont2
  A\kern-.1667em\lower.5ex\hbox{M}\kern-.125emS}}

\title{\sax Observations of 2 Jy Lobe-dominated Broad-Line Sources: \\
the Discovery of a Hard X-ray Component}

\author{Raffaella Morganti \address{Istituto di Radioastronomia,  Via Gobetti 101, 
I-40129 Bologna, Italy},
{Paolo Padovani} \address{Dipartimento di Fisica, II Universit\`a di Roma
``Tor Vergata'',  Via della Ricerca Scientifica 1, 
I-00133 Roma, Italy \\ Space Telescope Science Institute, 3700 San Martin 
Drive, Baltimore, MD. 21218, USA \\Affiliated to the Astrophysics Division,
Space Science Department, European Space Agency},
{Joachim Siebert} \address{Max-Planck Institut f\"ur Extraterrestrische
Physik, Giessenbachstrasse,  D-85740 Garching bei M\"unchen, Germany},
{Andrea Cimatti} \address{Osservatorio Astrofisico di Arcetri, Largo E. Fermi 5,
I-50125 Firenze, Italy},\\
{Clive N. Tadhunter} \address{Dept. of Physics, University of Sheffield, Sheffield
S3 7RH, UK}
and {Fausto Vagnetti} \address{Dipartimento di Fisica, II Universit\`a di Roma
``Tor Vergata'', Via della Ricerca Scientifica 1, I-00133 Roma, Italy}}

\begin{document}

\begin{abstract}
We present new {\it BeppoSAX} LECS, MECS, and PDS observations of five
lobe-dominated, broad-line active galactic nuclei selected from the 2 Jy
sample of southern radio sources. These include three radio quasars and two
broad-line radio galaxies. ROSAT PSPC data, available for all the objects, are
also used to better constrain the spectral shape in the soft X-ray band. The
collected data cover the $0.1 - 10$ keV energy range, reaching 40 keV for one
source. Detailed spectral fitting shows that all sources have a flat hard
X-ray spectrum with energy index $\alpha_{\rm x} \sim 0.75$ in the $2 - 10$ keV
energy range. This is a new result, which is at variance with the situation at
lower energies where these sources exhibit steeper spectra. Spectral breaks
$\sim 0.5$ at $1 - 2$ keV characterize the overall X-ray spectra of our
objects. The flat, high-energy slope is very similar to that displayed by
flat-spectrum/core-dominated quasars, which suggests that the same emission
mechanism (most likely inverse Compton) produces the hard X-ray spectra in
both classes. Contrary to the optical evidence for some of our sources, no
absorption above the Galactic value is found in our sample. Finally, a (weak)
thermal component is also present at low energies in the two broad-line radio
galaxies included in our study.
\end{abstract}

\maketitle

\begin{table*}
\setlength{\tabcolsep}{0.5pc}
\newlength{\digitwidth} \settowidth{\digitwidth}{\rm 0}
\catcode`?=\active \def?{\kern\digitwidth}
\caption{Sample Properties}
\begin{tabular}{lllcccc}
\hline
 Name   & $z$     & R$_{2.3GHz}$  & Class & LECS  &   MECS   & Observing   \\
        &         &               &       & exposure(s) & exposure(s) & date\\
\hline
\\
OF--109    & 0.574  & 0.57  & ~QSO & ...   & 16,864 & 22-Sep-96\\
Pictor A   & 0.035  & 0.03  & BLRG& ?5,271 & 14,798 & 12/13-Oct-96\\
OM--161  & 0.554  & 0.16  & ~QSO & 13,271 & 28,248 & 11/12-Jan-97 \\
PHL 1657   & 0.200  & 0.06  & ~QSO & ?4,704 & 17,429 & 29-Oct-96 \\
PKS 2152--69   & 0.027  & 0.03   & BLRG & ... &  ?9,367 & 29-Sep-96 \\
\hline
\end{tabular}
\end{table*}

\section{The Astrophysical Problem}

There is abundant evidence that strong anisotropies play a major role in the
observed characteristics of radio loud active galactic nuclei (AGN) and this
has led to suggest a unification of all high-power radio sources (Barthel
1989; see Antonucci 1993 and Urry \& Padovani 1995 for a review). 

According to this scheme:

1) the lobe-dominated, steep-spectrum radio quasars (SSRQ) and the
core-dominated, flat-spectrum radio quasars (FSRQ) are believed to be
increasingly aligned versions of Fanaroff-Riley type II (FR~II; Fanaroff \&
Riley 1974) radio galaxies. 

2) the broad-line (FWHM $\approx$ 2000~\kms) radio galaxies (BLRG) have a
still uncertain place.  They could represent either objects intermediate
between quasars and radio galaxies, (i.e.  with the nucleus only partly
obscured and the broad emission lines just becoming visible at the edge of the
obscuring torus) or low-redshift, low-power equivalent of quasars. 

As expected from this scheme, a nuclear, likely beamed X-ray component in
quasars is quite well established in particular for FSRQ and blazars (Wilkes
\& Elvis 1987; Shastri et al. 1993; Sambruna et al. 1994)

{\it What about the lobe-dominated quasars and broad-line radio galaxies?}

\section{The Sample}

We have observed with {\it BeppoSAX}  five lobe-dominated broad-line AGN
from the 2 Jy sample of radio sources (Wall \& Peacock 1985).  These have
been selected by taking the 10 sources (out of the 16 present in a
complete sub sample) with estimated flux in the 0.1 -- 10 keV band larger than
$2 \times 10^{-12}$ erg cm$^{-2}$ s$^{-1}$.  Unlike many previous studies of
the X-ray spectra of radio sources, our sample is well-defined and unbiased
(apart from the obvious limitations of all flux-limited samples).  The list of
objects and their characteristics are given in Table 1\footnote{Throughout
this paper the values $H_0 = 50$ km s$^{-1}$ Mpc$^{-1}$ and $q_0 = 0$ have
been adopted and spectral indices are written $S_{\nu} \propto
\nu^{-\alpha}$.}. 

Note that for these objects optical spectra (Tadhunter et al.  1993, 1997), radio
images and fluxes (Morganti et al.  1993, 1997) and ROSAT All-Sky Survey
and/or ROSAT PSPC pointed observations (Siebert et al.  1996) are available. 

\section{ Observations and Spectral Fits}

In the last three columns of Table 1 we summarize some information about the
observations

Data analysis was based on the linearized, cleaned event files obtained from
the {\it BeppoSAX} SDC on-line archive and on the XIMAGE package. Spectra
were accumulated for each observations using the SAXSELECT tool and the 
recommended extraction radii. 

Spectral analysis was performed with the XSPEC 9.00 package, using the response
matrices released by SDC in early 1997. The spectra were rebinned such that 
each new bin contains at least 20 counts.
Spectral fits for the three objects for which LECS data were available
show that the fitted N$_{\rm H}$ values are consistent with the
Galactic ones.  We then assumed Galactic N$_{\rm H}$ in the combined
LECS and MECS fits. For the two objects without LECS data this
assumption is also justified by the fact that the fit to the MECS data
is not strongly dependent on N$_{\rm H}$.

{}From the combined LECS and MECS fits all sources have very flat X-ray energy
indices, with a mean value $\langle \alpha_{\rm x} \rangle = 0.75\pm 0.02$. 
One object (PHL 1657) shows evidence of a spectral break $\sim 0.5$ at $E 
\sim 1$ keV. 

All our objects have ROSAT PSPC data (4 targets and one from the
RASS).  We then put the {\it BeppoSAX} and ROSAT PSPC data together to
better constrain the shape of the X-ray spectra, especially at low
energies. Two of our sources, in fact, have no LECS data, while for
the remaining three only less than 10 LECS bins are available below 1
keV.

At first, we fitted the {\it BeppoSAX} and ROSAT data separately with
a single power-law model modified at low energies by absorption due to
neutral gas in the Galaxy. From this preliminary analysis it was
clear that two different slopes for the power law are necessary at low
and high energies.

\begin{figure} 
{\psfig{figure=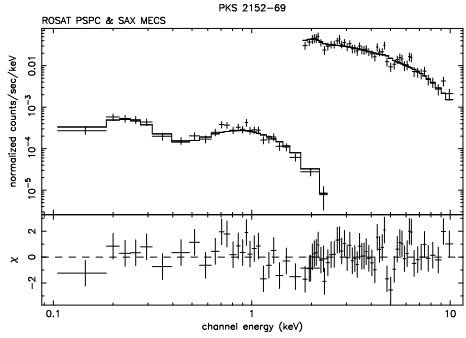,width=7cm,angle=-90}}
\caption{ROSAT PSPC and BeppoSAX MECS for the BLRG 2152-69 } 
\end{figure}

\begin{figure} 
{\psfig{figure=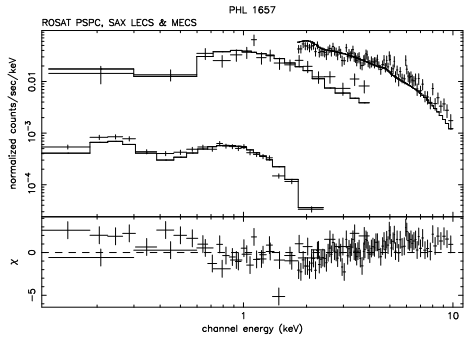,width=7cm,angle=-90}}
\caption{ROSAT PSPC, LECS and MECS for the SSRQ PHL 1657} 
\end{figure}

\begin{table*}
\setlength{\tabcolsep}{1.0pc}
\catcode`?=\active \def?{\kern\digitwidth}
\caption{ROSAT PSPC, BeppoSAX LECS and MECS spectral fits}
\begin{tabular}{lccccc}
\hline
 Name   & $\alpha_S$     & $\alpha_H$  & $E_{break}$ & $\chi^2_{\nu}$(dof)  &
Flux(2 -- 10 keV)   \\
        &                 &            &  keV        &     & 10$^{-12}$ erg cm$^{-2}$ s$^{-1}$ \\
\hline
\\
OF--109    & $1.33^{+0.11}_{-0.10}$  & $0.80^{+0.20}_{-0.21}$  
& $1.32^{+1.39}_{-0.37}$ &1.06(71)?& $?3.7\pm 0.1$ \\
Pictor A   & $0.79^{+0.07}_{-0.07}$  & $0.56^{+0.21}_{-0.38}$  
& $3.92^{+4.88}_{-3.10}$  &0.91(157)& $13.8 \pm 0.3$ \\
OM--161  & ...  & $0.77^{+0.11}_{-0.11}$  &  ... &0.79(86)?& 
$?2.5 \pm 0.1 $  \\
PHL 1657   & $1.42^{+0.09}_{-0.09}$  & $0.75^{+0.12}_{-0.12}$  &
$1.45^{+0.50}_{-0.29}$  &0.96(124)& $?8.6 \pm 0.2 $ \\
PKS 2152--69   & $1.23^{+0.57}_{-0.41}$  & $0.70^{+0.18}_{-0.18}$  & $1.69^{+2.01}_{-0.75}$  &0.98(71)?&  $?7.5\pm 0.2$ \\
\hline
\end{tabular}
\end{table*}

Therefore,  the data were merged together and fitted with a broken power
law.  The results are summarized in Table 2. The combined spectra for
the BLRG 2152-69 and the SSRQ PHL 1657 are shown in Figs.1 and 2 respectively.
(Note that the ROSAT counts [lower spectra]
have been normalized by the PSPC geometric area of 1,141 cm$^2$ and should be 
read as ``normalized counts/sec/keV/cm$^2$.'') 
The model parameters are extremely well
determined.  The spectra are obviously concave, with $\langle \alpha_{\rm S} -
\alpha_{\rm H} \rangle = 0.49 \pm 0.09$ and energy breaks around 1.5 keV
(Pictor A has a break at about 4 keV but with a large error due to the
relatively small difference between the soft and the hard spectral indices). 

One source, Pictor A, was also detected in the PDS up to $\approx 30 - 40$
keV.  The PDS appears to lie on the extrapolation of the LECS and MECS fit,
although given the relatively short PDS exposure time (6.8 ks) it is hard to
be more quantitative. 

The ROSAT PSPC images of our two lowest redshift objects, Pictor A and PKS
2152$-$69, show evidence for an extended component, probably related to the
diffuse emission from hot gas associated with early-type galaxies. Addition
of a thermal component to the power-law model results in a significant 
improvement of the fit, with best-fit gas temperatures $\sim 0.5$ keV. The
thermal component, however, contributes only $5 - 10\%$ to the total X-ray
flux, which is dominated by non-thermal (power-law) emission.

\begin{figure} 
{\psfig{figure=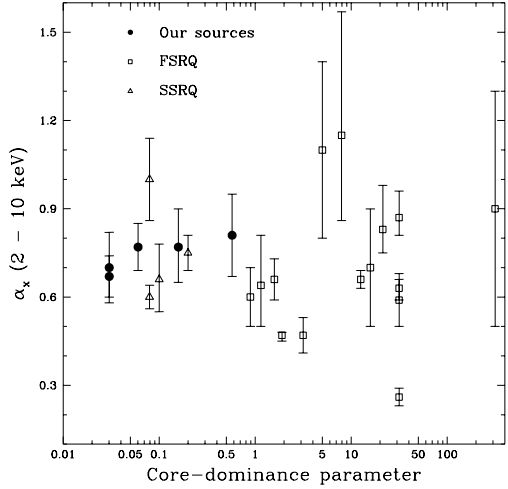,width=7cm,angle=0}}
\caption{$\alpha_X$ vs core dominance R for SSRQ and BLRG from our sample
(filled circles), SSRQ and FSRQ from literature (open triangles and squares
respectively)} 
\end{figure}

\section{Results}

Our main results are:

1)  lobe-dominated, broad-line AGN have a hard X-ray spectrum with
$\alpha_{\rm X} \sim 0.75$ at $E > 1 - 2$ keV similar to what found in
core-dominated radio quasars.  Fig. 3 shows the $2 - 10$ keV energy indices
for core dominated FSRQ and for the objects in our sample.  The FSRQ sample is
heterogeneous while our sample, although relatively small, is well defined and
has very well determined spectral indices.  Nevertheless, the FSRQ are
characterized by $\langle \alpha_{\rm 2-10 keV} \rangle = 0.70\pm0.06$,
consistent with our results. 

Thus, it seems natural to attribute the hard X-ray component to the same
emission mechanism suggested for FSRQ, i.e. synchrotron self-Compton 
emission.

2) various previous studies had found that SSRQ displayed a steep soft
X-ray spectrum.  In fact, despite the hard component at higher energies,  we
nevertheless observe a steeper spectrum at lower energies.  Our best fits to
the whole $0.1 - 10$ keV range indeed {\it require} a spectral break $\sim
0.5$ between the soft and hard energy slopes at about $1 - 2$ keV.  The
dispersion in the energy indices is larger for the soft component.  We find
$\sigma({\alpha{\rm S}}) = 0.28$ while $\sigma({\alpha{\rm H}}) = 0.10$, which
might suggest a more homogeneous mechanism at higher energies. 

This steep component could be related to the UV bump observed in most
radio-quiet quasars or be due to synchrotron emission  or a
combination of both. Work is in progress to tackle this problem.

3)  Finally, a thermal component is also present at low energies in the two
BLRG but contributes only $\approx 10\%$ of the flux.

\end{document}